\documentclass[preprint,preprintnumbers,amssymb,prd,nofootinbib,tightenlines]{revtex4}
\usepackage{graphicx}
\usepackage{axodraw}
\usepackage{dcolumn}
\usepackage{bm}

\begin{document}

\baselineskip=15pt
\parskip=5pt

\preprint{SLAC-PUB-9199}
\preprint{UK/TP-2002-05} 

\vspace*{0.5in}  

\title{Nonresonant Contributions in {\boldmath$B\to\rho\pi$}  Decay}

\author{Jusak Tandean${}^1$}
\email{jtandean@pa.uky.edu}

\author{S. Gardner${}^{1,2}$}   
\email{gardner@pa.uky.edu}

\affiliation{
${}^1$Department of Physics and Astronomy, University of Kentucky, 
Lexington, Kentucky 40506-0055\footnote{Permanent Address.}
}

\affiliation{
${}^2$Stanford Linear Accelerator Center, Stanford University, 
Stanford, California 94309
\vspace{5ex} \\  
}


\begin{abstract} 
We consider nonresonant contributions in the Dalitz-plot analysis 
of $B\to \rho\pi \to \pi^+\pi^- \pi^0$ decay and their potential 
impact on the extraction of the Cabibbo-Kobayashi-Maskawa 
 parameter $\alpha$.
In particular, we examine the role of the heavy mesons 
$B^*$  and  $B_0^{}$,  via the process  
$B\to \pi (B^*, B_0) \to  \pi^+\pi^-\pi^0$, 
and their interference with resonant contributions 
in the $\rho$-mass region. 
We discuss the inherent uncertainties and suggest that the 
effects may be substantially smaller than previously indicated. 
\end{abstract}

\pacs{}

\maketitle

\section{Introduction}   
   
The recent observation of $CP$  violation in the $B$-meson system, 
realized through the measurement of a nonzero, time-dependent, 
$CP$-violating asymmetry in the process   
$\,B^0\bigl(\bar B^0\bigr)\to J/\psi K_S^{}\,$ 
(and related ones)~\cite{expt}, heralds a new era of discovery.  
The result yields a value of $\sin(2\beta)$ in accord with 
standard model (SM) expectations~\cite{smx}, where  $\beta$,  
defined by  
$\,\exp({\rm i}\beta)\equiv -V_{cb}^*V_{cd}^{}/
\bigl(V_{tb}^*V_{td}^{}\bigr),\,$  
is an angle of the unitarity triangle,  
$V_{ij}^{}$ being an element of the Cabibbo-Kobayashi-Maskawa 
(CKM) matrix~\cite{ckm}.    
Ascertaining the presence of physics beyond the SM thus demands
the determination of all the angles of the unitarity triangle.

In this paper, we consider the decays  
$\,B^0 (\bar B^0) \to\rho\pi \to \pi^+\pi^-\pi^0,\,$ 
as a Dalitz-plot analysis of the possible $\rho\pi$ final 
states, under the assumption of isospin symmetry, permits 
the determination of the CKM parameter $\alpha$~\cite{qui&co}, 
where  $\,\alpha = \pi - \beta -\gamma\,$  and 
$\,\exp({\rm i}\gamma)\equiv -V_{ub}^* V_{ud}^{}/
\bigl(V_{cb}^*V_{cd}^{}\bigr).\,$ 
Our interest is in assessing the size of the nonresonant 
contributions which could possibly obscure the analysis, 
and in ameliorating their impact.  
Indeed, the strategy for the extraction of $\alpha$ relies, 
in part, on the assumption that the $\rho$ mesons dominate 
the $3\pi$ final state.   
There are, however, empirical indications that this assumption 
may not always be warranted. 
For example, combining the CLEO measurements of the branching 
fractions,  
$\,{\cal B}\bigl( \bar B^0\to\rho^\mp\pi^\pm \bigr) 
=\bigl(27.6_{-7.4}^{+8.4}\pm4.2\bigr)\times10^{-6}\,$  
and  
$\,{\cal B}\bigl( B^-\to\rho^0\pi^- \bigr) 
=\bigl(10.4_{-3.4}^{+3.3}\pm2.1\bigr)\times10^{-6}\,$~\cite{cleo},  
with the BABAR result 
$\,{\cal B}(B^0\to\rho^\pm\pi^\mp)=
(28.9\pm5.4\pm4.3)\times10^{-6}\,$~\cite{babar}  
yields
\begin{eqnarray}   \label{R_x}    
{\cal R}  \,=\,  
{ {\cal B}\bigl( \bar B^0\to\rho^\mp\pi^\pm \bigr)  \over  
 {\cal B}\bigl( B^-\to\rho^0\pi^- \bigr) }   
\,=\,  2.7\pm 1.2   \,\,,  
\end{eqnarray}     
where we have added the errors in quadrature and ignored correlations. 
These ratios are small~\cite{GaoWur} with respect to simple theoretical 
estimates, which give  $\,{\cal R}\sim 6 \,$~\cite{bsw}.   
An interesting possibility for the resolution of this  
discrepancy has been suggested in Refs.~\cite{dea&co,DeaPol}, 
whose authors investigate the possible backgrounds to   
$\,B\to\rho\pi\to3\pi\,$  decay which arise from contributions 
mediated by other resonances. 
They find that the light $\sigma$  resonance, a broad $I=J=0$
enhancement in $\pi\pi$ scattering, as well as 
the heavy-meson resonances  $B^*$   
$\bigl(J^P=1^-\bigr)$  and  $B_0^{}$  $\bigl(J^P=0^+\bigr)$, 
can modify the  $\,B\to3\pi$ branching ratios in the $\rho$-mass 
region and give rise to values of $\cal R$ crudely 
compatible with the empirical value of Eq.~(\ref{R_x}), 
given its large error. 
In particular, the contribution of $\,B^-\to\sigma\pi^-\,$ decay
significantly enhances the effective  $\,B^-\to\rho^0\pi^-\,$ 
branching ratio and lowers the value of  ${\cal R}$.   
Analogously, the $\sigma$  modestly impacts the 
$\,B^0\to\rho^0 \pi^0\,$  branching ratio~\cite{GarMei};
let us consider the issues.  
     
The analysis of $\,B^0 (\bar B^0) \to\rho\pi\to \pi^+\pi^-\pi^0\,$  
decay posits a two-step process, that is, that 
the amplitude for  $\,B^0\to\pi^+\pi^-\pi^0\,$ decay can be written as
\begin{equation}
A(B^0 \to \pi^+\pi^-\pi^0) \,=\, 
f_+^{} a_{+-}^{} + f_-^{} a_{-+}^{} + f_0^{} a_{00}^{} \,\,,
\end{equation}
where $\,a_{ij}^{}\equiv A(B^0\to \rho^i \pi^j)\,$
and  $f_i^{}$  is the vector form-factor describing 
$\,\rho^i\to\pi\pi\,$~\cite{qui&co}. 
An analogous construct can be made for 
$\,B^0 \bigl(\bar B^0\bigr) \to\sigma\pi\to \pi^+\pi^-\pi^0\,$ 
decay, which contains the scalar form-factor describing  
$\,\sigma \to\pi^+\pi^-.\,$  
It is evident that the manner in which the $\sigma$ populates
the $\rho$ phase-space will depend on the amplitude for 
$\,B^0\to \sigma \pi^0\,$ decay, as well as on the accompanying 
scalar form-factor. 
The $\sigma$ is a state of definite $CP$, so that the isospin 
analysis of Ref.~\cite{qui&co} can be enlarged to include
it~\cite{GarMei}; nevertheless, the analysis relies on the form
factors adopted for the $\,\rho^i\to\pi\pi\,$  and  
$\,\sigma\to\pi\pi\,$ processes.   
The resonances of interest are broad, so that Breit-Wigner
form-factors are generally insufficient: they do not satisfy 
general theoretical constraints, such as analyticity and unitarity,
over the $\pi\pi$ invariant-mass interval needed.
As discussed in detail in Ref.~\cite{GarMei}, the differences are 
striking for the scalar form-factor, and the resulting numerical 
impact on  $\,B\to3\pi\,$  decay is sizable.   
In contrast, the numerical differences for the vector form-factor 
are not large.

The purpose of this paper is to extend the work of  
Ref.~\cite{GarMei},  which deals exclusively with the $\rho$  
and  $\sigma$  contributions.  
We incorporate the  $B^*$  and  $B_0^{}$  contributions suggested 
in Ref.~\cite{dea&co}, as the effects they find in the 
$\,B^0\to\rho^0\pi^0\,$  channel are considerable.   
In this paper, however, we show that the off-shell nature of the 
$B^*$ and $B_0^{}$ weak and strong vertices adds considerably
to the uncertainty of the estimate of Ref.~\cite{dea&co} and
may well reduce these contributions significantly. 
Nevertheless, we also explore kinematical cuts which would be useful 
in reducing the impact of these effects in the $\rho$-mass region.

We begin in Sec.~\ref{hamiltonian} with the weak, effective 
Hamiltonian and the matrix elements pertinent to our calculations.  
Subsequently, in Sec.~\ref{amplitudes},  we derive the amplitudes 
associated with the various contributions of interest in
the $\rho$-mass region of $\,B \to 3\pi\,$  decay.
We discuss our numerical results in Sec.~\ref{results} and 
conclude in Sec.~\ref{conclusion}.

\section{Effective Hamiltonian and matrix elements\label{hamiltonian}} 
     
The effective, $|\Delta B|=1$ Hamiltonian for 
$\,b\to d q\bar q\,$  decay is given by~\cite{BucBL}      
\begin{eqnarray}   \label{H_eff}    
{\cal H}_{\rm eff}^{}  \,=\,  
{G_{\rm F}^{}\over\sqrt 2} \left[ 
\lambda_u^{} \left( C_1^{} O_1^{u} + C_2^{} O_2^{u} \right)   
+ \lambda_c^{} \left( C_1^{} O_1^{c} + C_2^{} O_2^{c} \right)   
- \lambda_t^{}\sum_{i=3}^{10} C_i^{} O_i^{} 
\right]   \,\,,  
\end{eqnarray}     
where $G_{\rm F}^{}$  is the Fermi coupling constant,  
$\,\lambda_q^{}\equiv V_{qb}^{} V_{qd}^{*}\,$ are CKM factors,   
$C_i^{}$  are Wilson coefficients, and  
$O_i^{}$ are four-quark operators.  
The expressions for  $C_i^{}$ and $O_i^{}$ are detailed in Ref.~\cite{BucBL}, 
though we interchange  
$\,C_1^{} O_1^q\leftrightarrow C_2^{} O_2^q\,$, 
so that  $\,C_1^{}\sim 1\,$  and  $\,C_1^{}>C_2^{}\,$.
We neglect the electroweak-penguin operators  $O_{7,\cdots,10}^{}$  
because their coefficients  $C_{7,\cdots,10}^{}$  
are smaller than the others.  
In the decay amplitudes that we derive,  the  $C_i^{}$  enter 
through the combinations  
$\,a_i^{}=C_i^{} + C_{i+1}^{}/N_{\rm c}^{}\,$  if  $i$  is odd 
and  $\,a_i^{}=C_i^{} + C_{i-1}^{}/N_{\rm c}^{}\,$  if  $i$  is even, 
where  $\,N_{\rm c}^{}=3\,$  is the number of colors.

\begin{figure}[t]         
\includegraphics{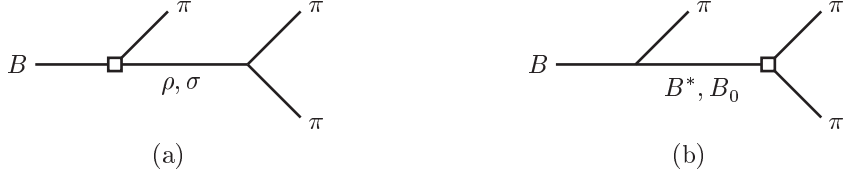}   \vspace{-3ex}  
\caption{\label{diagrams}   
Diagrams contributing to  $\,B\to3\pi,\,$ decay 
with each square denoting a weak vertex.}  
\end{figure}

The diagrams contributing to the  $\,B\to 3\pi\,$  amplitudes 
considered here, as shown in Fig.~\ref{diagrams}, each have a strong 
vertex and a weak vertex, where the latter describes 
the transition  
$\,M_b^{}\to M_1^{} M_2^{},\,$  in which  $M_b^{}$  is a heavy 
meson containing a $b$ quark and  $M_{1,2}^{}$  are light mesons.  
The amplitude corresponding to the weak vertex is given by  
\begin{eqnarray}   
A\bigl(M_b^{}\to M_1^{}M_2^{}\bigr)  \;=\;  
\bigl\langle M_1^{} M_2^{} \bigr| {\cal H}_{\rm eff}^{} 
\bigl| M_b^{} \bigr\rangle    \;.    
\end{eqnarray}     
To evaluate this, we adopt the naive factorization approximation, 
following earlier calculations~\cite{dea&co,DeaPol,GarMei} to
which we compare. 
   
The relevant matrix elements are      
\begin{eqnarray}   \label{0->pi}
\begin{array}{c}   \displaystyle      
\bigl\langle \pi^-(p) \bigl| \bar d \gamma^\mu L u 
\bigr| 0 \bigr\rangle  
\,=\,  
\sqrt 2\, \bigl\langle \pi^0(p) \bigl| \bar u \gamma^\mu L u 
\bigr| 0 \bigr\rangle  
\,=\,  {\rm i} f_\pi^{}\, p^\mu   \,\,,      
\vspace{1ex} \\   \displaystyle  
\bigl\langle \rho^- (p,\varepsilon) \bigr| \bar d \gamma^\mu u 
\bigl| 0 \bigr\rangle   
\,=\,  
\sqrt 2\, \bigl\langle \rho^0 (p,\varepsilon) \bigr| \bar u \gamma^\mu u 
\bigl| 0 \bigr\rangle   
\,=\,  f_\rho^{}\, \varepsilon^{*\mu}   \,\,,    
\end{array}   
\end{eqnarray}   
\begin{eqnarray}  \label{B->r,p,s} 
\begin{array}{c}   \displaystyle      
q^\mu \bigl\langle \rho^+(p,\varepsilon) \bigr| 
\bar u \gamma_\mu^{} L b
\bigl| \bar B^0(k) \bigr\rangle  
\,=\,  
-2{\rm i} A_0^{B\to\rho}(q^2) \, M_\rho^{}\, \varepsilon^*\cdot q    \,\,,  
\vspace{1ex} \\   \displaystyle  
\bigl\langle \pi^+(p) \bigr| \bar u \gamma^\mu L b
\bigl| \bar B^0 (k)\bigr\rangle  
\,=\,  
(k+p)^\mu\, F_1^{B\to\pi} (q^2) 
\,+\,  
{M_B^2-M_\pi^2\over q^2}\, q^\mu 
\left( F_0^{B\to\pi}(q^2)-F_1^{B\to\pi}(q^2) \right)   \,\,,  
\vspace{1ex} \\   \displaystyle  
q^\mu \bigl\langle \sigma(p) \bigr| 
\bar d\gamma_\mu^{}L b \bigl| \bar B^0(k) \bigr\rangle  
\,=\,   
-{\rm i} \left( M_B^2-M_\sigma^2 \right) F_0^{B\to\sigma}(q^2)  \,\,,  
\end{array}   
\end{eqnarray}   
\begin{eqnarray}   \label{B*,B0->p}    
\begin{array}{c}   \displaystyle      
q^\mu \bigl\langle \pi^+(p) \bigr| 
\bar u\, \gamma_\mu^{}L\, b
\bigl| \bar B^{*0}\bigl(k,\varepsilon_{B^*}^{}\bigr) \bigr\rangle  
\,=\,  
2{\rm i}\sqrt2\, A_0^{B^*\to\pi}(q^2)\, M_{B^*}^{}\, 
\varepsilon_{B^*}^{}\cdot q   \,\,,      
\vspace{1ex} \\   \displaystyle  
q^\mu \bigl\langle \pi^+(p) \bigr| 
\bar u \gamma^\mu L b \bigl| \bar B_0^0(k) \bigr\rangle  
\,=\,  
-{\rm i} \left( M_{B_0^{}}^2-M_\pi^2 \right) 
F_0^{B_0^{}\to\pi}(q^2) \,\,,     
\end{array}   
\end{eqnarray}   
where  $f_\pi^{}$  and  $f_\rho^{}$  are the usual decay constants, 
$q\equiv k - p$, and $L\equiv 1 - \gamma_5$. 
The various $A_0^{}(q^2)$ and $F_{0,1}^{}(q^2)$  are form factors.  
Other meson-to-meson matrix elements can be determined using 
isospin symmetry.  
In our phase convention,  the 
meson flavor wave functions are given by $\,\pi^+=u\bar d,\,$  
$\,\sqrt2\, \pi^0=u\bar u-d\bar d,\,$  $\pi^-=d\bar u,\,$  
$\,\bar B^0=b\bar d,\,$  $\,B^-=b\bar u,\,$  
and similarly for the  $\rho$,  $B^*$,  and  $B_0^{}$.  
This implies  that we have, for example, 
$\, \bigl\langle \pi^+ \bigr| \bar u\gamma_\mu^{}b
\bigl| \bar B^0 \bigr\rangle 
=-\sqrt 2\, \bigl\langle \pi^0 \bigr| 
\bar d\gamma_\mu^{} b \bigl| \bar B^0 \bigr\rangle 
=+\sqrt 2\, \bigl\langle \pi^0 \bigr| 
\bar u\gamma_\mu^{}b \bigl| B^- \bigr\rangle 
= \bigl\langle \pi^- \bigr| \bar d\gamma_\mu^{}b
\bigl| B^- \bigr\rangle .\,$  
We now employ these matrix
elements to realize amplitudes for $\,B\to 3\pi\,$ decays.

\section{Amplitudes\label{amplitudes}}    
    
Practical considerations drive our interest in the 
$\,\pi^+\pi^-\pi^0$  
and  $\pi^{\mp}\pi^{\mp}\pi^{\pm}\,$  decay modes; we
shall not consider  the  $\,\pi^0\pi^0\pi^\pm\,$ ones.     
We write the amplitude for   
$\,\bar B^0\to\pi^+\bigl(p_+^{}\bigr)\, \pi^-\bigl(p_-^{}\bigr)\, 
\pi^0\bigl(p_0^{}\bigr) \,$ decay as a coherent sum of the $\rho$, 
$\sigma$, $B^*$, and  $B_0^{}$ amplitudes, namely, 
\begin{eqnarray}    
\label{BZdecomp}
A^{+-0}  \,=\,  
A_\rho^{+-0} + A_\sigma^{+-0} + A_{B^*}^{+-0} + A_{B_0^{}}^{+-0}   \;.  
\end{eqnarray}     
For 
$\,B^-\to\pi^-\bigl(p_1^{}\bigr)\, \pi^-\bigl(p_2^{}\bigr)\, 
\pi^+\bigl(p_+^{}\bigr),\,$   
the amplitude $A^{--+}$  can be constructed in an analogous manner.

We consider first the 
$\,B\to\rho\pi\to 3\pi\,$ contributions,    
represented by the diagram denoted by ``$\rho$'' 
in  Fig.~\ref{diagrams}(a).
For each $\rho^i$ diagram and $3\pi$ state, 
the amplitude is written as a product of 
an amplitude for the  $\,B\to\rho^i\pi^j\,$  weak transition and   
a vertex function  $\Gamma_{\rho\pi\pi}^{}$ describing   
the  $\,\rho^i\to\pi\pi\,$  form factor. 
Were the $\rho$ a narrow resonance, the Breit-Wigner (BW) form 
\begin{eqnarray}   \label{GrppBW}
\Gamma_{\rho\pi\pi}^{\rm BW}(s)  \,=\,   
{g_\rho^{}\over s-M_\rho^2+{\rm i}\Gamma_\rho^{} M_\rho^{}}  \,\,  
\end{eqnarray}   
would suffice, where 
$\sqrt{s}$  is the invariant mass of the $2\pi$ system and
$g_\rho^{}$  is the $\,\rho\to\pi\pi\,$  coupling constant. 
However, since the $\rho$ is not narrow ---
its width is some $20\%$ of its mass --- 
this form must be generalized to accommodate known 
theoretical constraints over the region in $s$ for which
it is appreciable. 
For example, unitarity and time-reversal
invariance compel the phase of $\Gamma_{\rho\pi\pi}^{}(s)$ to be
that of $\,L=1$, $I=1\,$ $\pi\pi$ scattering for 
$\,s\lesssim \bigl(M_\pi^{} + M_\omega^{}\bigr)^2,\,$   
for which the scattering  is elastic. 
Moreover, the imaginary part of 
$\Gamma_{\rho\pi\pi}^{}(s)$ must vanish below physical threshold, 
$\,s= 4M_\pi^2.\,$   
For a detailed discussion with references to earlier
work, see Refs.~\cite{GarOCon1,GarMei}. 
Following Ref.~\cite{GarOCon1}, we have
\begin{eqnarray}   \label{Grpp}
\Gamma_{\rho\pi\pi}^{}(s)  \,=\,   
{-F_\rho^{}(s)\over f_{\rho\gamma}^{}}   \,\,,  
\end{eqnarray}   
where $F_\rho^{}(s)$  is the vector form-factor of the pion 
and  $f_{\rho\gamma}^{}$  is the  $\rho$-$\gamma$ coupling  
constant. 
The parameters in  $F_\rho^{}(s)$  are determined by 
fitting to  $\,e^+e^-\to\pi^+\pi^-\,$  data; what is important
is that the parametrization itself is consistent with 
theoretical constraints.  
The value of $f_{\rho\gamma}^{}$ is determined from the 
$\,\rho\to e^+e^-\,$  width, which, in turn, is extracted from 
the  $\,e^+e^- \to \pi^+\pi^-\,$  cross section at  
$\,s=M_\rho^2\,$~\cite{GarOCon1,GarOCon2}.    
The overall sign is chosen so that Eq.~(\ref{Grpp}) is equivalent 
to the BW form, Eq.~(\ref{GrppBW}), as $\,s\to M_\rho^2\,$. 
At $\,s=M_\rho^2\,$  the BW form is compatible with the various 
theoretical constraints. 
In our numerical analysis, we adopt the ``solution $B$'' fit of 
Ref.~\cite{GarOCon1} for $F_\rho^{}$, for which 
$\,f_{\rho\gamma}^{}=0.122 \pm 0.001\,\rm GeV^2\,$~\cite{GarOCon2}. 
Alternatively, a BW form with a running width  $\Gamma_\rho^{}(s)$, 
chosen to be compatible with the form of the $\pi\pi$ phase 
shift (in the crossed channel) as  $\,s\to 4 M_\pi^2,\,$    
is given in Ref.~\cite{babarbook}. 
However, the numerical differences between this form and 
the one we have chosen are small~\cite{GarMei}.

For the decay amplitudes, after summing over the $\rho$  
polarizations, we find  
\begin{eqnarray}   \label{A_rho}  
\begin{array}{c}   \displaystyle   
A_\rho^{+-0}  \,=\,  
\eta^-\, \bigl(s_{+-}^{} -  s_{+0}^{}\bigr) \,  
\Gamma_{\rho\pi\pi}^{} \bigl( s_{-0}^{} \bigr) 
\,+\,  
\eta^+\, \bigl( s_{-0}^{}  - s_{+-}^{}\bigr) \,  
\Gamma_{\rho\pi\pi}^{} \bigl( s_{+0}^{} \bigr)   
\,-\,    
\eta^0\, \bigl( s_{-0}^{}-s_{+0}^{} \bigr) \,   
\Gamma_{\rho\pi\pi}^{} \bigl( s_{+-}^{} \bigr)   \,\,,   
\vspace{2ex} \\   \displaystyle   
A_\rho^{--+}  \,=\,  
-\bar\eta^0\, \Bigl[ 
\bigl( s_{12}^{}-s_{1+}^{} \bigr) \, 
\Gamma_{\rho\pi\pi}^{} \bigl( s_{2+}^{} \bigr) 
\,+\,  
\bigl( s_{12}^{}-s_{2+}^{} \bigr) \,   
\Gamma_{\rho\pi\pi}^{} \bigl( s_{1+}^{} \bigr) 
\Bigr]   \,\,,  
\end{array}     
\end{eqnarray}     
where   
$\,s_{kl}^{}\equiv \bigl( p_k^{}+p_l^{} \bigr) ^2,\,$  
with 
\begin{eqnarray}    
\begin{array}{c}   \displaystyle   
\eta^-  \,=\,  
{G_{\rm F}^{}\over\sqrt 2} \left(  
\lambda_u^{}\, a_1^{} - \lambda_t^{}\, a_4^{}    
\right) f_\rho^{}\, F_1^{B\pi}   \,\,,   
\hspace{2em}  
\eta^+  \,=\,     
{G_{\rm F}^{}\over \sqrt 2} \left[   
\lambda_u^{}\, a_1^{}   
- \lambda_t^{} \left( a_4^{} - a_6^{}\, R_q^{} \right)   
\right] f_\pi^{} M_\rho^{}\, A_0^{B\rho}   \,\,,  
\vspace{2ex} \\   \displaystyle   
\eta^0  \,=\,    
{-G_{\rm F}^{}\over 2\sqrt 2} \left\{ 
\left[   
\lambda_u^{}\, a_2^{}+\lambda_t^{} \left(a_4^{}-a_6^{}\, R_q^{}\right)  
\right] f_\pi^{} M_\rho^{}\, A_0^{B\rho}     
\,+\,  
\left( \lambda_u^{}\, a_2^{} + \lambda_t^{}\, a_4^{} \right) 
f_\rho^{}\, F_1^{B\pi}   
\right\}   \,\,,   
\vspace{2ex} \\   \displaystyle   
\bar\eta^0  \,=\,   
{G_{\rm F}^{}\over 2} \left\{  
\left[ 
\lambda_u^{}\, a_1^{}-\lambda_t^{} \left(a_4^{}-a_6^{}\, R_q^{}\right) 
\right] f_\pi^{} M_\rho^{}\, A_0^{B\rho}       
\,+\,   
\left( \lambda_u^{}\, a_2^{} + \lambda_t^{}\, a_4^{} \right) 
f_\rho^{}\, F_1^{B\pi} 
\right\}   \,\,.    
\end{array}     
\end{eqnarray}     
Here  $\,A_0^{B\rho}\equiv A_0^{B\to\rho} \bigl( M_\pi^2 \bigr)\,$   
and  $\,F_1^{B\pi}\equiv F_1^{B\to\pi} \bigl(M_\rho^2 \bigr),\,$  
whereas  
$\,R_q^{}\equiv 
M_\pi^2/\bigl[\bigl(m_b^{}+\hat{m}^{}\bigr)\hat{m}^{}]\,$ --- note that
we work in the isospin-symmetric limit, for which 
$\,\hat{m}^{}=m_u^{}=m_d^{}\,$. 
The relative signs between the different terms in Eq.~(\ref{A_rho})  
follow from  the $\rho\pi\pi$  couplings\footnote{
We use the notation 
$\,\bigl\langle M_2^{} M_3^{}\bigr|M_1^{}\bigr\rangle\equiv
\bigl\langle M_2^{} M_3^{}\bigr|{\cal H}_{\rm strong}^{}
\bigr|M_1^{}\bigr\rangle.\,$}    
\begin{eqnarray}   \label{M(rho->2pi)}  
\begin{array}{c}   \displaystyle   
\bigl\langle\pi^0\bigl(p_0^{}\bigr)\,   
\pi^\pm\bigl(p_\pm^{}\bigr)\big| \rho^\pm\bigr\rangle  
\,=\,  
\pm g_\rho^{}\, 
\varepsilon_\rho^{}\cdot \bigl(p_\pm^{}-p_0^{}\bigr)   \,\,,   
\vspace{2ex} \\   \displaystyle   
\bigl\langle\pi^+\bigl(p_+^{}\bigr)\,   
\pi^-\bigl(p_-^{}\bigr) \big| \rho^0\bigr\rangle 
\,=\,      
g_\rho^{}\, 
\varepsilon_\rho^{}\cdot \bigl( p_-^{}-p_+^{} \bigr)   \,\,,   
\end{array}   
\end{eqnarray}   
which follow, in turn, from the phase conventions we have
chosen for the flavor wave functions: 
$\,\bigl|\pi^\pm \bigr\rangle=\mp \bigl|I=1,I_3=\pm 1\bigr\rangle\,$
and  $\,\bigl|\pi^0 \bigr\rangle = \bigl|I=1,I_3=0 \bigr\rangle,\,$  
and similarly for the~$\rho$ states.   
Our  $A_\rho^{}$  amplitudes agree with those of earlier 
calculations~\cite{ali,dea&co,GarMei}.

We turn next to the $\sigma$ ``meson'' contributions, represented 
by the diagram denoted by ``$\sigma$'' in Fig.~\ref{diagrams}(a). 
We use the $\sigma$ to denote a two-pion state with total isospin  
$\,I=0\,$  and total angular-momentum  $\,J=0\,$; it need not
be a ``pre-existing'' resonance, but, rather, can be 
generated dynamically by the strong pionic final-state 
interactions in this channel~\cite{OO}. 
The peak of the broad enhancement associated with the  $\sigma$ 
is close to the $\rho$  in mass, so that the decay  
$\,B\to\sigma\pi\to 3\pi\,$  can populate the  $\,B\to\rho\pi\,$  
phase space~\cite{DeaPol}.   
As in the $\rho$ case, the amplitudes for  
$\,B\to\sigma\pi\to 3\pi\,$ decays are written as a product of 
an amplitude for  the $\,B\to\sigma\pi$ weak transition 
and a vertex function  $\Gamma_{\sigma\pi\pi}^{}$  describing 
the  $\,\sigma\to\pi\pi\,$  form factor. 
We write~\cite{GarMei}   
\begin{eqnarray}   \label{Gspp}  
\Gamma_{\sigma\pi\pi}^{}(s)  \,=\,  \chi\, \Gamma_1^{n*}(s)   \,\,,  
\end{eqnarray}   
where  $\Gamma_1^n$  is defined as   
\begin{eqnarray}   \label{G1n}
\bigl\langle0\bigr| \bar d d\bigl|\pi^+(p_+)\pi^-(p_-)\bigr\rangle  
\,=\,  
\sqrt{2\over3}\, {\Gamma_1^n(s_{+-})\, B_0^{} }   \,\,. 
\end{eqnarray}   
We note that  $\,B_0^{}\equiv M_\pi^2/(2\hat{m})\,$  is 
the vacuum quark condensate and $\chi$  is a normalization 
constant, to be discussed shortly. 
For our numerical work in the next section, we 
adopt the  $\Gamma_1^n(s)$ as derived in Ref.~\cite{MeiOll}, 
after Refs.~\cite{npa,basde,OO}.   
The calculated form factor is realized in a chiral, unitarized, 
coupled-channel approach; at low energies, the form factor is 
matched to the one-loop-order expression in  
chiral perturbation theory~\cite{GasLeu,MeiOll}. 
The resulting form factor is consistent with low-energy 
constraints and is comparable to the scalar form-factor which 
emerges from the dispersion analysis of Ref.~\cite{DGL};
however, it is notably different from the Breit-Wigner form 
adopted in Refs.~\cite{e791,DeaPol} to study the role of the  
$\sigma$  in $B$ and $D$ decays into the $3\pi$ final state.  
That is, 
\begin{eqnarray}   \label{GsppBW}  
\Gamma_{\sigma\pi\pi}^{\rm BW}(s)  \;=\;  
{ g_{\sigma\pi\pi}^{}  \over  
 s-M_\sigma^2+{\rm i}^{}\Gamma_\sigma^{}(s)\, M_\sigma^{} }  \,\,,  \,\,\,  
\quad 
\Gamma_\sigma^{}(s)=\frac{M_\sigma \Gamma_\sigma^{}}{\sqrt{s}}
\sqrt{s-4M_\pi^2\over M_\sigma^2 -4 M_\pi^2}   \;,
\end{eqnarray}   
where the coupling 
$\,g_{\sigma\pi\pi}^{}\equiv\langle\pi^+\pi^-| 
\sigma\rangle\,$  
is determined from the  $\,\sigma\to\pi\pi\,$  decay 
rate.
For $\,B\to3\pi\,$ decay, the numerical changes arising from 
the use of  $\Gamma_{\sigma\pi\pi}^{}(s)$  
in place of the BW expression are significant~\cite{GarMei}, 
as we will see here as well. 
We determine the normalization $\chi$ by 
requiring that~\cite{GarMei}   
\begin{eqnarray}   \label{spp}
\chi \left| \Gamma_1^n \bigl(M_\sigma^2\bigr) \right|  \,=\,  
{ g_{\sigma\pi\pi}^{}   \over  
 \Gamma_\sigma^{}\bigl(M_\sigma^2\bigr) \, M_\sigma^{} }   \,\,,  
\end{eqnarray}   
which equates $\bigl|\Gamma_{\sigma\pi\pi}(s)\bigr|$  to its BW 
counterpart at  $\,s=M_\sigma^2.\,$     
The values of  $M_\sigma$  and  $\Gamma_\sigma$  
are extracted from fits of  $\Gamma_{\sigma\pi\pi}^{\rm BW}(s)$ to 
$\,D\to3\pi\,$  decays~\cite{e791}.    
The normalization condition is motivated by noting that the modulus 
of  $\Gamma_1^n(s)$  is peaked near  $\,s=M_\sigma^2,\,$  whereas
the normalization of $\Gamma_1^n(s)$ is sensitive to the 
values of certain, poorly known low-energy constants~\cite{GarMei}.
We emphasize that $M_\sigma^{}$  and  $\Gamma_\sigma^{}$  appear
merely in the normalization of $\Gamma_{\sigma\pi\pi}^{}$.

The resulting decay amplitudes are then   
\begin{eqnarray}    
A_\sigma^{+-0}  \,=\,   
\eta_\sigma^0\, \Gamma_{\sigma\pi\pi}^{} \bigl(s_{+-}^{}\bigr)   \,\,,  
\hspace{2em}  
A_\sigma^{--+}  \,=\,  
\bar\eta_\sigma^0 \left( 
\Gamma_{\sigma\pi\pi}^{} \bigl(s_{1+}^{}\bigr) 
+ \Gamma_{\sigma\pi\pi}^{} \bigl(s_{2+}^{}\bigr) 
\right)   \,\,,  
\end{eqnarray}     
where   
\begin{eqnarray}    
\begin{array}{c}   \displaystyle   
\eta_\sigma^0  \,=\,  
{G_{\rm F}^{}\over 2} \left\{ \left[   
\lambda_u^{}\, a_2^{}   
+ \lambda_t^{} \left( a_4^{}-a_6^{}\, R_q^{} \right)  
\right] \left( M_B^2-M_\sigma^2 \right) f_\pi^{} F_0^{B\sigma}
\,-\,   
\lambda_t^{}\, a_6^{}\, 
{2 \langle\sigma|\bar d d|0\rangle\over m_b^{}-\hat{m}^{}} 
\left( M_B^2-M_\pi^2 \right) F_0^{B\pi} 
\right\}  ,   
\vspace{2ex} \\   \displaystyle   
\bar\eta_\sigma^0  \,=\,  
{G_{\rm F}^{}\over\sqrt 2} \left\{ \left[  
\lambda_u^{}\, a_1^{}   
- \lambda_t^{} \left( a_4^{}-a_6^{}\, R_q^{} \right) 
\right] \left( M_B^2-M_\sigma^2 \right) f_\pi^{} F_0^{B\sigma}   
\,+\,   
\lambda_t^{}\, a_6^{}\, 
{2 \langle\sigma|\bar d d|0\rangle\over m_b^{}-\hat{m}^{}} 
\left( M_B^2-M_\pi^2 \right) F_0^{B\pi}   
\right\}   ,    
\end{array}     
\end{eqnarray}     
with  
$\,F_0^{B\pi}\equiv F_0^{B\to\pi} \bigl(M_\sigma^2\bigr)\,$  and  
$\,F_0^{B\sigma}\equiv F_0^{B\to\sigma}\bigl(M_\pi^2\bigr).\,$ 
From  Eqs.~(\ref{Gspp}) and~(\ref{G1n}), it follows that  
$\,\langle\sigma|\bar d d|0\rangle= 
M_\pi^2/\bigl(\sqrt 6\,\chi \hat{m}^{}\bigr).\,$   
We agree with the weak amplitudes of Ref.~\cite{GarMei}, 
but disagree with those of Ref.~\cite{DeaPol} in that our  
$\eta_\sigma^0$  and  $\bar\eta_\sigma^0,\,$ 
neglecting penguin terms, are smaller and larger, 
respectively, than theirs by a factor of $\sqrt 2$.

We now evaluate the  $B^*$  and  $B_0^{}$  contributions, 
whose diagrams are shown in Fig.~\ref{diagrams}(b); we suppose
that other excited $B$-meson states could also contribute, 
but we expect that their larger masses ought to make them less 
important~\cite{dea&co}.
Presently, no reliable data exist on the widths of these 
heavy mesons, so that their values have to be calculated.
Recent estimates~\cite{B-review,BecLey} suggest that the $B^*$  
is a very narrow resonance, whereas the  $B_0^{}$  is less so, 
its width being some $6\%$ of its mass.    
Nevertheless, the resonances are sufficiently narrow
that it is reasonable to adopt a Breit-Wigner representation 
for the propagators of these mesons, as in Ref.~\cite{dea&co}.  
In the combined heavy-quark and chiral limit~\cite{hq}, 
the strong couplings connecting the  $\bigl(B^*,B_0^{}\bigr)$,  
$B$, and $\pi$ mesons are~\cite{dea&co,B-review,B0}\footnote{
We note that  
$\,\langle B^{*-}(k,\varepsilon)\pi^+|\bar B^0\rangle =
-\langle B^-\pi^+|\bar B^{*0}(k,\varepsilon)\rangle\,$  
and  
$\,\langle B_0^-(k)\pi^+|\bar B^0\rangle  
=-\langle B^-\pi^+|\bar B_0^0(k)\rangle.\,$  
}   
\begin{equation}   \label{BsBp} 
\bigl\langle B^-\bigl(p^\prime\bigr)\pi^+(p) 
\bigr| \bar B^{*0}(k,\varepsilon) \bigr\rangle  
\,=\,   
-{2 g\, \sqrt{M_B^{} M_{B^*}^{}}\over f_\pi^{}}\, 
\varepsilon\cdot p^{}   \,\,,  
\end{equation}
\begin{equation} \label{B0Bp} 
\bigl\langle B^-(p^\prime)\pi^+(p) \bigr| \bar B_0^0(k) \bigr\rangle  \,=\,   
{ h\, \sqrt{M_B^{} M_{B_0^{}}^{}}\over f_\pi^{}}\, 
{k^2-M_B^2\over M_{B_0^{}}^{}}   \;.
\end{equation}
Using isospin symmetry, we derive 
\begin{eqnarray}    
\bigl\langle B^- \pi^+ \bigr| \bar B^{*0}\bigr\rangle  
\,=\,  
-\sqrt 2\, \bigl\langle \bar B^0\pi^0 \bigr|  \bar B^{*0} \bigr\rangle  
\,=\,  \bigl\langle \bar B^0 \pi^- \bigr| B^{*-} \bigr\rangle   \,\,  
\end{eqnarray}   
and analogous relations for  
$\bigl\langle B\pi \bigr| B_0^{} \bigr\rangle$.  
We then obtain       
\begin{eqnarray}   \label{A_B*} 
\begin{array}{c}   \displaystyle   
A_{B^*}^{+-0}  \;=\;   
{ \frac{1}{\sqrt 2} K\, \Pi \bigl(s_{-0}^{},s_{+-}^{}\bigr) 
 + K_1^{}\, \Pi \bigl(s_{-0}^{},s_{+0}^{}\bigr)
 \over  s_{-0}^{}-M_{B^*}^2+{\rm i}\Gamma_{B^*}^{}M_{B^*}^{} }  
\,-\,   
{ \frac{1}{\sqrt 2} K\, \Pi \bigl(s_{+-}^{},s_{-0}^{}\bigr)
 \over  s_{+-}^{}-M_{B^*}^2+{\rm i}\Gamma_{B^*}^{}M_{B^*}^{} }   \,\,,  
\vspace{2ex} \\   \displaystyle   
A_{B^*}^{--+}  \,=\,  
{ K\, \Pi \bigl(s_{1+}^{},s_{12}^{}\bigr)
 \over  s_{1+}^{}-M_{B^*}^2+{\rm i}\Gamma_{B^*}^{}M_{B^*}^{} }   
\,+\,  
{ K\, \Pi \bigl(s_{2+}^{},s_{12}^{}\bigr)   
 \over  s_{2+}^{}-M_{B^*}^2+{\rm i}\Gamma_{B^*}^{}M_{B^*}^{} }  \,\,,  
\end{array}   
\end{eqnarray}   
\begin{eqnarray}   \label{A_B0}
\begin{array}{c}   \displaystyle   
A_{B_0^{}}^{+-0}  \,=\,  
\left(  
{ \tilde K^0 + \tilde K^{cc}  \over 
 s_{-0}^{}-M_{B_0^{}}^2+{\rm i}\Gamma_{B_0^{}}^{} M_{B_0^{}}^{} }  
\,-\,  
{ \tilde K^0 \over  
 s_{+-}^{}-M_{B_0^{}}^2+{\rm i}\Gamma_{B_0^{}}^{} M_{B_0^{}}^{} }  
\right) \frac{\left(M_{B_0^{}}^2-M_\pi^2\right)}{\sqrt 2}   \,\,,  
\vspace{2ex} \\   \displaystyle   
A_{B_0^{}}^{--+}  \,=\,  
\left(  
{ \tilde K^0  \over 
 s_{1+}^{}-M_{B_0^{}}^2+{\rm i}\Gamma_{B_0^{}}^{} M_{B_0^{}}^{} }  
\,+\,  
{ \tilde K^0  \over   
 s_{2+}^{}-M_{B_0^{}}^2+{\rm i}\Gamma_{B_0^{}}^{} M_{B_0^{}}^{} }  
\right) \left( M_{B_0^{}}^2-M_\pi^2 \right)   \,\,,  
\end{array}   
\end{eqnarray}   
where   
\begin{eqnarray}    
\begin{array}{c}   \displaystyle   
K  \,=\,   
-4 G_{\rm F}^{} \left[     
\lambda_u^{}\, a_1^{}-\lambda_t^{} \left(a_4^{}-a_6^{}\, R_q^{}\right)  
\right] g M_{B^*}^{}\sqrt{M_B^{} M_{B^*}^{}}\, A_0^{B^*\pi}   \,\,,  
\vspace{2ex} \\   \displaystyle   
K_1^{}  \,=\,   
-2\sqrt 2\, G_{\rm F}^{} \left[     
\lambda_u^{}\, a_2^{}+\lambda_t^{} \left(a_4^{}-a_6^{}\, R_q^{}\right)   
\right] g M_{B^*}^{}\sqrt{M_B^{} M_{B^*}^{}}\, A_0^{B^*\pi}   \,\,,  
\end{array}     
\end{eqnarray}     
\begin{eqnarray}    
\begin{array}{c}   \displaystyle   
\tilde K^0  \,=\,   
{G_{\rm F}^{}\over\sqrt 2} \left[  
\lambda_u^{}\, a_1^{}   
- \lambda_t^{} \left(a_4^{}-a_6^{}\, R_q^{}\right)  
\right] {M_{B_0^{}}^2-M_B^2\over M_{B_0^{}}^{}}\,   
h\, \sqrt{M_B^{} M_{B_0^{}}^{}}\, F_0^{B_0^{}\pi}   \,\,,  
\vspace{2ex} \\   \displaystyle   
\tilde K^{cc}  \,=\,  
{G_{\rm F}^{}\over\sqrt 2} \left[  
\lambda_u^{}\, a_2^{}   
+ \lambda_t^{} \left(a_4^{}-a_6^{}\, R_q^{}\right)  
\right]  {M_{B_0^{}}^2-M_B^2\over M_{B_0^{}}^{}}\,   
\, \sqrt{M_B^{} M_{B_0^{}}^{}}\, F_0^{B_0^{}\pi}   \,\,,  
\end{array}     
\end{eqnarray}     
and the sum over $B^*$ polarizations yields 
\begin{eqnarray}    
\Pi(u,v)  \,=\,     
{ \bigl( M_B^2-M_\pi^2-u \bigr) \, u\over 4M_{B^*}^2} 
\,+\,  M_\pi^2 - {v\over2}   \,\,.    
\end{eqnarray}     
Note that 
$\,A_0^{B^*\pi}\equiv A_0^{B^*\to\pi}\bigl(M_\pi^2\bigr)\,$  and  
$\,F_0^{B_0^{}\pi}\equiv F_0^{B_0^{}\to\pi}\bigl( M_\pi^2 \bigr).\,$
Our expressions for  $A^{+-0}$  in Eqs.~(\ref{A_B*}) and~(\ref{A_B0}) 
disagree with those in Ref.~\cite{dea&co} in that the factors 
of $1/\sqrt 2$  are missing in their formulas, and that the minus 
sign in the middle of the big brackets in Eq.~(\ref{A_B0}) is  
opposite to theirs.  
However, our expressions for  $A^{--+}$  in Eqs.~(\ref{A_B*}) 
and~(\ref{A_B0}) agree with theirs.

\section{Numerical results and discussion\label{results}}    
     
We begin by listing the parameters that we use; 
we conform with the parameter choices of Refs.~\cite{dea&co,GarMei},
in order to realize a crisp comparison with their results. 
In specific, the Wilson coefficients we use are  
\begin{eqnarray}    
\begin{array}{c}   \displaystyle   
C_1^{}  \,=\,  1.100   \,,  \hspace{0.5em}  
C_2^{} \,=\,  -0.226   \,,  \hspace{0.5em}  
C_3^{} \,=\,   0.012   \,,  \hspace{0.5em}  
C_4^{} \,=\,  -0.029   \,,  \hspace{0.5em}  
C_5^{} \,=\,   0.009   \,,  \hspace{0.5em}  
C_6^{} \,=\,  -0.033   \,.         
\end{array}      
\end{eqnarray}      
For the CKM factors, we adopt the Wolfenstein 
parametrization~\cite{wolfenstein}, retaining terms 
of ${\cal O}\bigl(\lambda^3\bigr)$  in the real part and  
of ${\cal O}\bigl(\lambda^5\bigr)$  in the imaginary part, to wit, 
\begin{eqnarray}    
\begin{array}{c}   \displaystyle   
V_{ud}^{}  \,=\,  1-\lambda^2/2   \,\,,  
\hspace{0.5em}  
V_{ub}^{}  \,=\,  A \lambda^3\, \bigl[\rho-{\rm i}\eta\, 
\bigl(1-\lambda^2/2\bigr)\bigr]   \,\,,  
\hspace{0.5em}  
V_{td}^{}  \,=\,  A \lambda^3\, (1-\rho-{\rm i}\eta)   \,\,,  
\hspace{0.5em}  
V_{tb}^{}  \,=\,  1  \,\,,   
\end{array}      
\end{eqnarray}      
and using 
\begin{eqnarray}    
\lambda  \,=\,  0.2196   \,\,,  \hspace{2em}  
\rho  \,=\,  0.05    \,\,,  \hspace{2em}  
\eta  \,=\,  0.36    \,\,,  \hspace{2em}  
A     \,=\,  0.806  \,\,.     
\end{eqnarray}      
For decay constants, light meson masses, and resonance parameters,
we have 
\begin{eqnarray}   
\begin{array}{c}   \displaystyle   
f_\pi^{}/\sqrt2  \,=\,  92.4\,{\rm MeV}   \,\,,   \hspace{2em} 
M_\pi^{} \,=\,  139.57\,{\rm MeV}   \,\,,   \hspace{2em} 
\vspace{1ex} \\   \displaystyle   
f_\rho^{}  \,=\,  0.15\,{\rm GeV}^2  \,\,,   \hspace{2em}  
M_\rho^{}  \,=\,  769.3\,{\rm MeV}  \,\,,   \hspace{2em}     
\Gamma_\rho^{}  \,=\,  150\,{\rm MeV}  \,\,,   \hspace{2em}     
g_\rho^{}  \,=\,  5.8   \,\,,   
\vspace{1ex} \\   \displaystyle   
M_\sigma^{}  \,=\,  478\,{\rm MeV}  \,\,,   \hspace{2em}     
\Gamma_\sigma^{}  \,=\,  324\,{\rm MeV}  \,\,,   \hspace{2em}     
g_{\sigma\pi\pi}^{}  \,=\,  2.52\,{\rm GeV}   \,\,.
\end{array}      
\end{eqnarray}      
The decay constants $f_\pi^{}$ and $f_\rho^{}$ are associated
with $\pi^\pm$ and $\rho^\pm$ decay, respectively. 
We neglect isospin-violating effects throughout, so that 
$\,M_{\pi^\pm}^{}  =  M_{\pi^0}^{}  = M_\pi^{},\,$ 
$\,M_{\rho^\pm}^{}  =  M_{\rho^0}^{}  = M_\rho^{},\,$ as well 
as  $\,M_{\bar B^0}^{}  =  M_{B^-}^{} =  M_B^{}.\,$ 
Moreover, $\,\hat{m}^{} =6\,{\rm MeV}.\,$    
The $B^*$  and $B$  are degenerate in the heavy-quark limit,  
so that we neglect their mass difference as well. 
We also neglect the lifetime difference between the $\bar B^0$ 
and $B^-$, setting $\,\tau_{\bar B^0}^{}=\tau_{B^-}^{}=\tau_B^{}.\,$ 
For the $B$ and related mesons, we have
\begin{eqnarray}   
\begin{array}{c}   \displaystyle   
M_B^{}  \,=\,  5.279\,{\rm GeV}   \,\,,   
\hspace{2em}     
\tau_B^{} = 1.6\times 10^{-12}\,{\rm s}  \,\,,   
\hspace{2em}     
m_b^{}  \,=\,  4.6\,{\rm GeV}   \,\,,   
\vspace{1ex} \\   \displaystyle   
\Gamma_{B^*}^{}  \,=\,  0.2\,{\rm keV}   \,\,,   \hspace{2em}     
M_{B_0^{}}^{}  \,=\,  5.697\,{\rm GeV}   \,\,,   \hspace{2em}   
\Gamma_{B_0^{}}^{}  \,=\,  0.36\,{\rm GeV}   \,\,,   \hspace{2em}   
\end{array}      
\end{eqnarray}      
and use  
\begin{eqnarray}   \label{g,h}
g  \,=\,  0.6   \,\,,  \hspace{3em}     
h  \,=\,  -0.7   \,\,.  
\end{eqnarray}      
The heavy-to-light transition form factors are given by
\begin{eqnarray}   
\begin{array}{c}   \displaystyle   
A_0^{B\rho}  \,=\,  0.29   \,\,,  
\hspace{2em}  
F_0^{B\pi}  \,=\,  0.37   \,\,,  \hspace{2em}  
F_1^{B\pi}  \,=\,  0.37   \,\,,  \hspace{2em}  
F_0^{B\sigma}  \,=\,  0.46   \,\,,  
\vspace{1ex} \\   \displaystyle   
A_0^{B^*\pi}  \,=\,  0.16   \,\,,  \hspace{2em}  
F_0^{B_0^{}\pi}  \,=\,  -0.19   \,\,.  \hspace{2em}  
\end{array}     
\end{eqnarray}     
Finally, for the vector and scalar form-factors, 
$\Gamma_{\rho\pi\pi}^{}(s)$  and  $\Gamma_{\sigma\pi\pi}^{}(s)$,
respectively,    
we follow the treatment of Ref.~\cite{GarMei}.  
The  $F_\rho^{}(s)$  parametrization we adopt was fit to 
$\,e^+e^-\to\pi^+\pi^-\,$  data in the elastic region~\cite{GarOCon1},  
$\,2 M_\pi^{}\le\sqrt{s}\le M_\pi^{}+M_\omega,\,$ only, so that
for larger values of $s$ we use a 
Breit-Wigner form, matched to the value of 
$\Gamma_{\rho\pi\pi}^{}(s)$  at $\,\sqrt s=M_\pi^{}+M_\omega.\,$
That is, for  $\,\sqrt s\gtrsim 923\,\rm MeV\,$  we employ 
$\,\Gamma_{\rho\pi\pi}^{}(s)=\bigl[ 
c_{\rm r}^{}\, \bigl( M_\rho^2-s \bigr) + 
{\rm i} c_{\rm i}^{}\Gamma_\rho^{} M_\rho^{} \bigr] g_\rho^{}/
\bigl[ \bigl( M_\rho^2-s \bigr)^2+\Gamma_\rho^2 M_\rho^2 \bigr] ,\,$  
with   $\,c_{\rm r}^{}\simeq 0.929\,$  and  
$\,c_{\rm i}^{}\simeq1.29.\,$    
For the scalar form-factor, we employ the
$\Gamma_1^n(s)$  derived in  Ref.~\cite{MeiOll},  
which is valid for  $\,\sqrt s\lesssim 1.2\,\rm GeV.\,$    
The normalization of  Eq.~(\ref{spp}) implies that 
$\,\chi=20.0\,{\rm GeV}^{-1}.\,$    
For $\,\sqrt s>1.2\,\rm GeV,\,$  we match to the asymptotic 
form of $\Gamma_{\sigma\pi\pi}^{}(s)\,$~\cite{DGL},  
as detailed in Ref.~\cite{GarMei}.

To obtain branching ratios for $\,B\to 3\pi\,$ decay in 
the $\rho$-mass region,  we integrate over the 
region of phase space satisfying the requirement 
that two of the three pions reconstruct the $\rho$ mass within an 
interval of 2$\delta$, as was done in Refs.~\cite{dea&co,GarMei}.   
This amounts in each case to calculating the effective width
\begin{eqnarray}   
\begin{array}{c}   \displaystyle   
\label{Geff}
\Gamma_{\rm eff}^{}\bigl(B\to \rho(p_1+p_2)\pi(p_3)\bigr)  
\,=\,  
\Gamma\bigl(B\to \pi(p_1) \pi(p_2) \pi(p_3)\bigr) \Bigl|_{
(M_\rho^{}-\delta)^2\le s_{12}^{}\le (M_\rho^{}+\delta)^2}^{}   \,\,.  
\end{array}     
\end{eqnarray}     
We choose  $\,\delta=0.3\,\rm GeV,\,$ 
following earlier work~\cite{dea&co,GarMei}.

For crisp comparison with Ref.~\cite{dea&co}, we begin by computing 
the effective branching ratios arising from the use of Breit-Wigner
forms, as in Eqs.~(\ref{GrppBW}) and~(\ref{GsppBW}) for the 
$\rho$ and $\sigma$, respectively, throughout.   
The various contributions, reflective of the enumerated terms 
in Eq.~(\ref{BZdecomp}), are reported in Table~\ref{0.3BW-P-CC}.  
There are differences between our results for 
the  $\rho$,  $\,\rho$$+$$B^*\,$,  and  
$\,\rho$$+$$B^*$$+$$B_0^{}\,$  contributions and 
the corresponding ones in Ref.~\cite{dea&co}.  
The differences are, however, not large and arise in part from 
missing factors in the formulas for the $B^*$  and  
$B_0^{}$ amplitudes,  which we delineated in the last section.  
In contrast, as pointed out in Ref.~\cite{GarMei},  
the  $\sigma$  effect on the $B^-$  decay is much bigger   
than that found in Ref.~\cite{DeaPol}, because our $\sigma$ 
amplitude is larger than theirs by a factor of $\sqrt2$.  
This is evident in the  $\,\rho$$+$$\sigma\,$  and  
$\,\rho$$+$$\sigma$$+$$B^*\,$  columns.   
Our results agree with those Ref.~\cite{GarMei}, to the extent
that they are applicable; we note that Ref.~\cite{GarMei}  
neglects penguin contributions altogether and deals 
exclusively with the  $\rho$  and  $\sigma$  contributions.
The last column of  Table~\ref{0.3BW-P-CC}  contains the 
sum of all the contributions,  
$\,\rho$$+$$\sigma$$+$$B^*$$+$$B_0^{}\,$.
Overall, it is apparent that the effect of the $B_0^{}$ is smaller 
than that of the other contributions, although it is not negligible.  
Finally, in the last row, we collect the ratios  of  
branching ratios ${\cal R}$ defined in Eq.~(\ref{R_x}).  
These  results show that the inclusion of the 
$\sigma$ and $B^*$, either individually or together, makes 
the estimated value of $\cal R$ consistent with the empirical one, 
given its large error.

\begin{table}[t]   
\caption{\label{0.3BW-P-CC}%
Effective branching ratios for  $\,B\to\rho\pi\,$  decays,  
as per Eq.~(\protect{\ref{Geff}}), with  $\,\delta=0.3\,\rm GeV.\,$  
Breit-Wigner form factors are used throughout, noting
Eqs.~(\protect{\ref{GrppBW}}) and (\protect{\ref{GsppBW}})
for the $\rho$  and  $\sigma$  contributions, respectively. 
All branching ratios are reported in units of $10^{-6}$.  
} \centering   \footnotesize
\vskip 0.5\baselineskip
\begin{tabular}
{@{\hspace{1ex}}c@{\hspace{1ex}}||@{\hspace{1ex}}c@{\hspace{1ex}}  
|@{\hspace{1ex}}c@{\hspace{1ex}}|@{\hspace{1ex}}c@{\hspace{1ex}}
|@{\hspace{1ex}}c@{\hspace{1ex}}|@{\hspace{1ex}}c@{\hspace{1ex}}
|@{\hspace{1ex}}c@{\hspace{1ex}}|@{\hspace{1ex}}c@{\hspace{1ex}}
|@{\hspace{1ex}}c@{\hspace{1ex}}|@{\hspace{1ex}}c@{\hspace{1ex}}}
\hline \hline  
Decay mode  &  $\vphantom{\biggl|}\rho$  &  $\sigma$  &  $B^*$  &  
$B_0$  &  $\rho+B^*$  &  $\rho+B^*+B_0^{}$  &  $\rho+\sigma$  &  
$\rho+\sigma+B^*$  &  $\rho+\sigma+B^*+B_0^{}$    
\\ \hline && && && && \vspace{-2ex} \\   
$\begin{array}{rcl}   \displaystyle
\bar B^0  & \hspace{-0.5em}
\rightarrow & \hspace{-0.5em} \rho^-\pi^+ \\     
\bar B^0  & \hspace{-0.5em} \rightarrow & \hspace{-0.5em} \rho^+\pi^- \\     
\bar B^0  & \hspace{-0.5em} \rightarrow & \hspace{-0.5em} \rho^0\pi^0 
\vspace{1.0ex}\\  
B^-  & \hspace{-0.5em} \rightarrow & \hspace{-0.5em} \rho^0\pi^-  
 \vspace{0.8ex}\end{array}$   
&  
$\begin{array}{c}   \displaystyle   
16.0  \\  4.76 \\  0.91 \vspace{1.0ex}\\ 4.10  
  \vspace{0.8ex}\end{array}$  
&  
$\begin{array}{c}   \displaystyle   
0.0003 \\ 0.0003 \\  0.045 \vspace{1.0ex}\\  5.18 
  \vspace{0.8ex}\end{array}$  
&  
$\begin{array}{c}   \displaystyle   
0.54  \\  0.13 \\  0.39 \vspace{1.0ex}\\ 2.71 
  \vspace{0.8ex}\end{array}$  
&  
$\begin{array}{c}   \displaystyle   
0.009  \\  0.020 \\  0.016 \vspace{1.0ex}\\ 0.107
  \vspace{0.8ex}\end{array}$  
&  
$\begin{array}{c}   \displaystyle   
16.5  \\ 4.98  \\ 1.43 \vspace{1.0ex}\\ 7.42   
  \vspace{0.8ex}\end{array}$  
&  
$\begin{array}{c}   \displaystyle   
16.3 \\ 4.98 \\ 1.29 \vspace{1.0ex}\\ 8.45 
  \vspace{0.8ex}\end{array}$  
&  
$\begin{array}{c}   \displaystyle   
16.0 \\ 4.78 \\ 0.93 \vspace{1.0ex}\\ 8.83 
  \vspace{0.8ex}\end{array}$  
&  
$\begin{array}{c}   \displaystyle   
16.4 \\ 5.00 \\ 1.59 \vspace{1.0ex}\\ 7.67 
  \vspace{0.8ex}\end{array}$  
&  
$\begin{array}{c}   \displaystyle   
16.3 \\ 5.00 \\ 1.43 \vspace{1.0ex}\\ 7.92 
  \vspace{0.8ex}\end{array}$  
\\ \hline \hline && && && && \vspace{-2ex} \\   
$\vphantom{\Bigl|}\cal R$
&  
5.1  
&  
- 
&  
- 
&  
- 
&  
2.9  
&  
2.5  
&  
2.3  
&  
2.8  
&  
2.7  
\\  \hline \hline  
\end{tabular}
\vspace{1ex} \\  
\end{table}

We now proceed to compute the effective branching ratios with
the $\rho$ and $\sigma$ form-factors, 
Eqs.~(\ref{Grpp}) and~(\ref{Gspp}),  which we advocate.  
These results are presented in Table~\ref{0.3-P-CC}. 
The results without the $\sigma$ contributions change little, 
as the vector form-factor is not terribly different from its BW  
counterpart~\cite{GarMei}.    
In the presence of the $\sigma$, this similarity persists for 
the $\bar B^0$  decays,  but, in contrast, the $B^-$  branching  
ratios are significantly increased compared to the corresponding  
ones in  Table~\ref{0.3BW-P-CC}.  
This effect also tends to diminish the
relative impact of the $B^*$  and $B_0$ contributions on the 
$\rho^0\pi^-$ mode, though the heavy mesons persist in making
a substantial impact on the effective branching ratio for the
$\rho^0\pi^0$ mode.

\begin{table}[b]   
\caption{\label{0.3-P-CC}%
Effective branching ratios for  $\,B\to\rho\pi\,$  decays,  
as per Eq.~(\protect{\ref{Geff}}), with  $\,\delta=0.3\,\rm GeV.\,$  
We adopt the $\rho$ and $\sigma$ form factors, 
Eqs.~(\protect{\ref{Grpp}}) and (\protect{\ref{Gspp}}), 
respectively, which we have advocated. 
All branching ratios are reported in units of $10^{-6}$.  
}  \centering   \footnotesize
\vskip 0.5\baselineskip
\begin{tabular}{@{\hspace{1ex}}c@{\hspace{1ex}}|  
|@{\hspace{1ex}}c@{\hspace{1ex}}
|@{\hspace{1ex}}c@{\hspace{1ex}}|@{\hspace{1ex}}c@{\hspace{1ex}}
|@{\hspace{1ex}}c@{\hspace{1ex}}|@{\hspace{1ex}}c@{\hspace{1ex}}
|@{\hspace{1ex}}c@{\hspace{1ex}}|@{\hspace{1ex}}c@{\hspace{1ex}}
|@{\hspace{1ex}}c@{\hspace{1ex}}|@{\hspace{1ex}}c@{\hspace{1ex}}}
\hline \hline  
Decay mode  &  $\vphantom{\biggl|}\rho$  &  $\sigma$  &  $B^*$  &  $B_0$  &  
$\rho+B^*$  &  $\rho+B^*+B_0^{}$  &  $\rho+\sigma$  &  
$\rho+\sigma+B^*$  &  $\rho+\sigma+B^*+B_0^{}$    
\\ \hline && && && && & \vspace{-2ex} \\   
$\begin{array}{rcl}   \displaystyle
\bar B^0  & \hspace{-0.5em} \rightarrow & \hspace{-0.5em} \rho^-\pi^+ \\     
\bar B^0  & \hspace{-0.5em} \rightarrow & \hspace{-0.5em} \rho^+\pi^- \\     
\bar B^0  & \hspace{-0.5em} \rightarrow & \hspace{-0.5em} \rho^0\pi^0  
\vspace{1.0ex}\\  
B^-  & \hspace{-0.5em} \rightarrow & \hspace{-0.5em} \rho^0\pi^-  
 \vspace{0.8ex}\end{array}$   
&  
$\begin{array}{c}   \displaystyle   
16.0  \\  4.76 \\  0.86 \vspace{1.0ex}\\ 4.06 
  \vspace{0.8ex}\end{array}$  
&  
$\begin{array}{c}   \displaystyle   
0.001  \\  0.001 \\  0.065 \vspace{1.0ex}\\ 7.66 
  \vspace{0.8ex}\end{array}$  
&  
$\begin{array}{c}   \displaystyle   
0.54  \\  0.13 \\  0.39 \vspace{1.0ex}\\ 2.71 
  \vspace{0.8ex}\end{array}$  
&  
$\begin{array}{c}   \displaystyle   
0.009  \\  0.020 \\  0.016 \vspace{1.0ex}\\ 0.107 
  \vspace{0.8ex}\end{array}$  
&  
$\begin{array}{c}   \displaystyle   
16.6 \\ 4.90 \\ 1.35 \vspace{1.0ex}\\ 7.20 
  \vspace{0.8ex}\end{array}$  
&  
$\begin{array}{c}   \displaystyle   
16.4 \\ 4.93 \\ 1.21 \vspace{1.0ex}\\ 8.25 
  \vspace{0.8ex}\end{array}$  
&  
$\begin{array}{c}   \displaystyle   
15.9 \\ 4.80 \\ 0.91 \vspace{1.0ex}\\ 11.1 
  \vspace{0.8ex}\end{array}$  
&  
$\begin{array}{c}   \displaystyle   
16.5 \\ 4.94 \\ 1.47 \vspace{1.0ex}\\ 11.9 
  \vspace{0.8ex}\end{array}$  
&  
$\begin{array}{c}   \displaystyle   
16.3 \\ 4.98 \\ 1.33 \vspace{1.0ex}\\ 12.7 
  \vspace{0.8ex}\end{array}$  
\\ \hline \hline && && && && & \vspace{-2ex} \\   
$\vphantom{\Bigl|}\cal R$
&  
5.1  
&  
- &  
- &  
- &  
3.0  
&  
2.6  
&  
1.9  
&  
1.8  
&  
1.7    
\\  \hline \hline    
\end{tabular}    
\vspace{1ex} \\ 
\end{table}   

\begin{figure}[ht]
\includegraphics{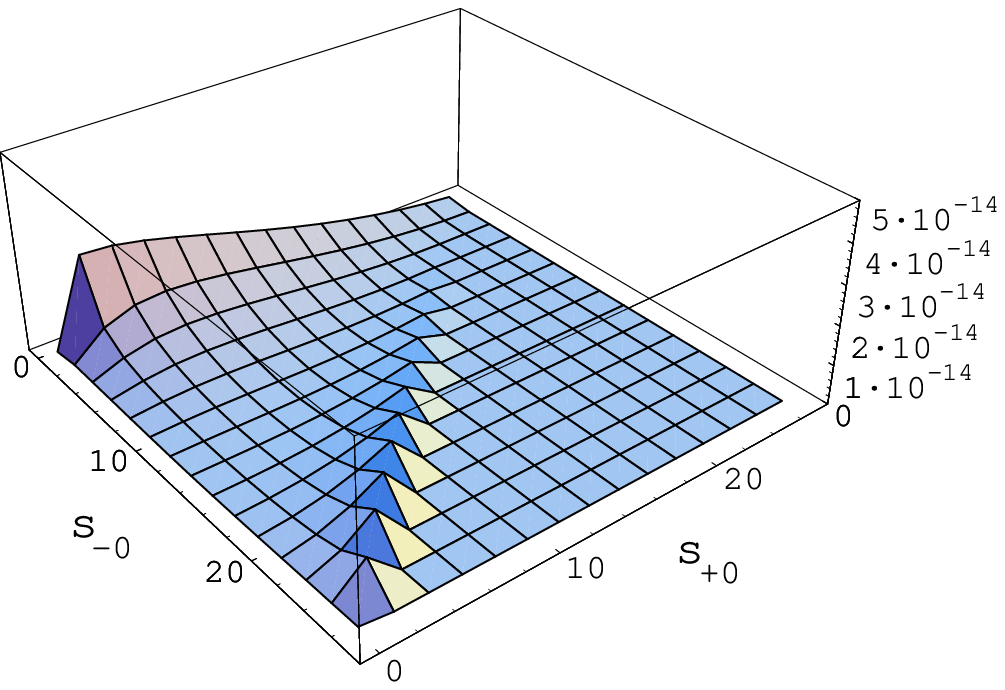}   \vspace{-3ex}
\caption{\label{fig:BZ3D} 
The $B^*$ and $B_0$ contributions to 
$\,\bar B^0 \to \pi^+(p_+)\pi^-(p_-)\pi^0(p_0)\,$ decay, specifically,
$\bigl|A_{B^*}^{+-0} + A_{B_0}^{+-0}\bigr|^2$  
(in dimensionless units)  as a function of its arguments 
$s_{+0}^{}$ and $s_{-0}^{}$, both in units of GeV$^2$. 
}   
%
%
\includegraphics{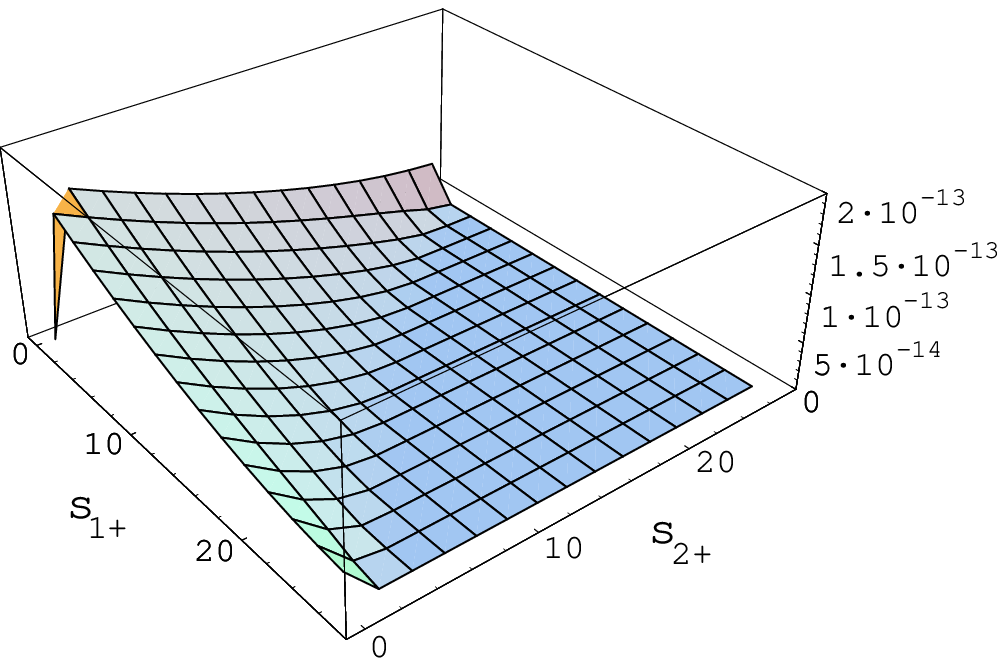}   \vspace{-4ex}
\caption{\label{fig:BM3D}
The $B^*$ and $B_0$ contributions to 
$\,B^- \to \pi^-(p_1)\pi^-(p_2)\pi^+(p_+)\,$ decay, specifically,
$\bigl|A_{B^*}^{--+} + A_{B_0}^{--+}\bigr|^2$  (in dimensionless 
units)  as a function of its arguments
$s_{1+}$ and $s_{2+}$, both in units of GeV$^2$. 
}
\end{figure}

Were the heavy-meson contributions to the  $\rho^0\pi^0$  
mode seen in Table~\ref{0.3-P-CC} as large as we have
estimated, the impact on the Dalitz-plot analysis to 
extract $\alpha$ from $\bar B^0 (B^0) \to \pi^+\pi^-\pi^0$ decays
would be significant~\cite{dea&co}.
Since the  $B^*$  and  $B_0^{}$  masses lie outside the phase-space 
region of  $\,B\to3\pi,\,$  their effects behave as part of 
the nonresonant background, but are not uniform and obviously 
interfere with other contributions. The manner in which the
contributions are distributed throughout the Dalitz plot is shown 
for  $\,\bar B^0\to \pi^+\pi^-\pi^0\,$ decay in 
Fig.~\ref{fig:BZ3D}; the heavy-meson contributions preferentially 
populate the edges of the Dalitz plot, in which the $\rho$ 
contributions lie as well.    
In $B^-\to \pi^+\pi^-\pi^-$ decay,  the distribution of the 
heavy-meson contributions is somewhat more uniform, 
as illustrated in Fig.~\ref{fig:BM3D}. 

We now proceed to consider the reliability of the estimates
we have effected. 
Let us first note that the parameters $g$ and $h$ of 
the strong heavy-meson couplings in Eqs.~(\ref{BsBp}), 
(\ref{B0Bp}) assume the values given in 
Eq.~(\ref{g,h}) --- these reflect the upper limits of their estimated 
ranges~\cite{dea&co,B-review}.\footnote{   
We also note, however,  that the  $g$  value in Eq.~(\ref{g,h}) 
is, by virtue of heavy-quark symmetry, favored by the recent 
measurement of the  $\,D^*\to D\pi\,$  width~\cite{D*width}, 
which yields  $\,g=0.59\pm 0.01\pm 0.07.\,$}
Thus, the results we find with these parameters 
can be regarded as extremal estimates
(although variations in other numerical inputs, such as 
the form factors, do introduce further uncertainties).  
Choosing central values of $g$ and $h$ in their estimated
ranges decreases the heavy-meson effects by up to some 
50$\%\,$~\cite{dea&co},
as explicitly shown in Table~\ref{0.3-med-P-CC}.

%
%
\begin{table}[ht]   
\caption{\label{0.3-med-P-CC}%
Effective branching ratios for  $\,B\to\rho\pi\,$  decays,  
as in Table~\protect{\ref{0.3-P-CC}}, except that 
$\,g=0.40\,$  and  $\,h=-0.54\,$  have been used.    
}  \centering   \footnotesize
\vskip 0.5\baselineskip
\begin{tabular}{@{\hspace{1ex}}c@{\hspace{1ex}}|  
|@{\hspace{1ex}}c@{\hspace{1ex}}|@{\hspace{1ex}}c@{\hspace{1ex}}
|@{\hspace{1ex}}c@{\hspace{1ex}}|@{\hspace{1ex}}c@{\hspace{1ex}}
|@{\hspace{1ex}}c@{\hspace{1ex}}|@{\hspace{1ex}}c@{\hspace{1ex}}
|@{\hspace{1ex}}c@{\hspace{1ex}}|@{\hspace{1ex}}c@{\hspace{1ex}}}
\hline \hline  
Decay mode  &  $\vphantom{\biggl|}\rho$  &  $B^*$  &  $B_0^{}$  &  
$\rho+B^*$  &  $\rho+B^*+B_0^{}$  &  $\rho+\sigma$  &  
$\rho+\sigma+B^*$  &  $\rho+\sigma+B^*+B_0^{}$    
\\ \hline && && && && \vspace{-2ex} \\   
$\begin{array}{rcl}   \displaystyle
\bar B^0  & \hspace{-0.5em} \rightarrow & \hspace{-0.5em} \rho^-\pi^+ \\     
\bar B^0  & \hspace{-0.5em} \rightarrow & \hspace{-0.5em} \rho^+\pi^- \\     
\bar B^0  & \hspace{-0.5em} \rightarrow & \hspace{-0.5em} \rho^0\pi^0  
\vspace{1.0ex}\\  
B^-  & \hspace{-0.5em} \rightarrow & \hspace{-0.5em} \rho^0\pi^-  
 \vspace{0.8ex}\end{array}$   
&  
$\begin{array}{c}   \displaystyle   
16.0  \\  4.76 \\  0.86 \vspace{1.0ex}\\ 4.06 
  \vspace{0.8ex}\end{array}$  
&  
$\begin{array}{c}   \displaystyle   
0.24  \\  0.06 \\  0.17 \vspace{1.0ex}\\ 1.20 
  \vspace{0.8ex}\end{array}$  
&  
$\begin{array}{c}   \displaystyle   
0.005  \\  0.012 \\  0.009 \vspace{1.0ex}\\ 0.064 
  \vspace{0.8ex}\end{array}$  
&  
$\begin{array}{c}   \displaystyle   
16.3 \\ 4.82 \\ 1.10 \vspace{1.0ex}\\ 5.55 
  \vspace{0.8ex}\end{array}$  
&  
$\begin{array}{c}   \displaystyle   
16.1 \\ 4.87 \\ 1.03 \vspace{1.0ex}\\ 6.12 
  \vspace{0.8ex}\end{array}$  
&  
$\begin{array}{c}   \displaystyle   
15.9 \\ 4.80 \\ 0.91 \vspace{1.0ex}\\ 11.1 
  \vspace{0.8ex}\end{array}$  
&  
$\begin{array}{c}   \displaystyle   
16.2 \\ 4.86 \\ 1.20 \vspace{1.0ex}\\ 11.0 
  \vspace{0.8ex}\end{array}$  
&  
$\begin{array}{c}   \displaystyle   
16.1 \\ 4.91 \\ 1.13 \vspace{1.0ex}\\ 11.4 
  \vspace{0.8ex}\end{array}$  
\\ \hline \hline && && && && \vspace{-2ex} \\   
$\vphantom{\Bigl|}\cal R$
&  
5.1  
&  
- 
&  
- 
&  
3.8  
&  
3.4  
&  
1.9  
&  
1.9  
&  
1.8    
\\   \hline \hline    
\end{tabular}    
\vspace{1ex} \\  
\end{table}   

Moreover, the relative signs chosen for the $\rho$,  $\sigma$,  
and  heavy-meson contributions will impact the numerical 
values of the effective branching ratios.   
As noted by Ref.~\cite{dea&co}, the relative sign of the 
$B^*$ and $B_0^{}$  contributions is fixed in the heavy-quark 
and chiral limits.  
The relative signs of the heavy-meson,  $\rho$,  and  $\sigma$ 
contributions, however, are less clear.   
We define the  $\rho\to\pi\pi$  coupling as per 
Eq.~(\ref{M(rho->2pi)}),  after Ref.~\cite{dea&co,GarMei}, 
though we note that a chiral Lagrangian analysis suggests that 
the relations of Eq.~(\ref{M(rho->2pi)}) should possess an 
additional overall sign.   
With this modification, the branching ratios for the  
$\,\rho$+$B^*$+$B_0^{}\,$  combination in Table~\ref{0.3-P-CC}
typically become smaller by no more than~15$\%$.  
However, the $\,\rho$+$\sigma\,$ results in 
$\,\bar B^0\to\rho^0\pi^0\,$  and  $\,B^-\to\rho^0\pi^-\,$
become some~3$\%$  and~10$\%$ larger, respectively.  
The impact on  the $\,\rho$+$\sigma$+$B^*$+$B_0^{}\,$  
results is mixed, leading to a suppression of about~10$\%$  
in  the $\rho^0\pi^0$ mode and an enhancement of~2$\%$   
in the $\rho^0\pi^-$ mode.

Kinematical cuts can mitigate the impact of the heavy-meson
and $\sigma$ contributions. 
Since the $\rho^\pm\pi^\mp$  modes are little affected by these
notions, we evaluate only the  $\rho^0\pi^0$  and  $\rho^0\pi^-$  
modes.   
We try two different sets of kinematical cuts.  
For the first one, we set  
$\,\delta=0.15\,{\rm GeV}=\Gamma_\rho^{}\,$  and report 
our results in Table~\ref{0.15-P-CC}.   
The relative suppression of
the heavy-meson and $\sigma$ contributions is quite modest,
if it exists at all. 
For the second set, we impose not only a  $\delta$  
cut but also a cut on $\cos\theta$,  where  $\theta$ is 
the helicity angle, defined as the angle between the direction 
of one member of a pion pair from $\rho$ decay and the direction 
of the parent $B$-meson evaluated in the pair's rest-frame.  
Since the $\rho$ contribution has a  $\cos^2$$\theta$ 
distribution in $B\to\pi^+\pi^-\pi^0$ 
decay~\cite{qui&co}, larger  values of $|\cos\theta|$  
enhance the $\rho$ contribution. Interference effects in 
the $B^-\to \pi^+\pi^-\pi^-$ channel will make this cut 
less effective. 
We set  $\,\delta=0.3\,\rm GeV\,$ and  $\,|\cos\theta|>0.4,\,$  
and collect the results in Table~\ref{0.3,0.4,P,CC}.   
Comparing to Table~\ref{0.3-P-CC}, the $\theta$ 
cut is seen to decrease the relative size of the $\sigma$ 
background, as discussed in Ref.~\cite{GarMei}.  
The helicity-angle cut only modestly reduces  the $\sigma$ 
contribution in  $\,B^-\to \rho^0\pi^-\,$  decay;  
however, an assumption of  $\rho$  dominance is only needed 
when one employs an isospin analysis to extract $\sin(2\alpha)$
from  $\,B^0 \bigl(\bar B^0 \bigr) \to \pi^+\pi^-\pi^0\,$  decay, 
as detailed in Ref.~\cite{qui&co}. 
For the  $\rho^0\pi^0$  mode, the chosen cut does significantly 
reduce an already small contribution. 
Were the $\sigma$ contribution to the $\rho^0 \pi^0$ mode 
much larger than we estimate, then a full partial-wave 
analysis to separate the $s$- and $p$-wave contributions
could be both practicable and necessary.

\begin{table}[t]   
\caption{\label{0.15-P-CC}%
Effective branching ratios for  $\,B\to\rho\pi\,$  decays,  
as in Table~\protect{\ref{0.3-P-CC}}, except that 
$\,\delta=0.15\,\rm GeV\,$  has been used. 
}  \centering   \footnotesize
\vskip 0.5\baselineskip
\begin{tabular}{@{\hspace{1ex}}c@{\hspace{1ex}}|  
|@{\hspace{1ex}}c@{\hspace{1ex}}|@{\hspace{1ex}}c@{\hspace{1ex}}
|@{\hspace{1ex}}c@{\hspace{1ex}}|@{\hspace{1ex}}c@{\hspace{1ex}}
|@{\hspace{1ex}}c@{\hspace{1ex}}|@{\hspace{1ex}}c@{\hspace{1ex}}
|@{\hspace{1ex}}c@{\hspace{1ex}}|@{\hspace{1ex}}c@{\hspace{1ex}}}
\hline \hline  
Decay mode  &  $\vphantom{\biggl|}\rho$  &  $\sigma$  &  $B^*$  &  
$\rho+B^*$  &  $\rho+B^*+B_0^{}$  &  $\rho+\sigma$  &  
$\rho+\sigma+B^*$  &  $\rho+\sigma+B^*+B_0^{}$    
\\ \hline && && && && \vspace{-2ex} \\   
$\begin{array}{rcl}   \displaystyle
\bar B^0  & \hspace{-0.5em} \rightarrow & \hspace{-0.5em} \rho^0\pi^0 
\vspace{1.0ex}\\  
B^-  & \hspace{-0.5em} \rightarrow & \hspace{-0.5em} \rho^0\pi^-  
 \vspace{0.8ex}\end{array}$   
&  
$\begin{array}{c}   \displaystyle   
0.33 \vspace{1.0ex}\\ 3.36   
   \vspace{0.8ex}\end{array}$  
&  
$\begin{array}{c}   \displaystyle   
0.029 \vspace{1.0ex}\\ 3.46      
   \vspace{0.8ex}\end{array}$  
&  
$\begin{array}{c}   \displaystyle   
0.20 \vspace{1.0ex}\\ 1.38  
   \vspace{0.8ex}\end{array}$  
&  
$\begin{array}{c}   \displaystyle   
0.55 \vspace{1.0ex}\\ 4.82    
    \vspace{0.8ex}\end{array}$  
&  
$\begin{array}{c}   \displaystyle   
0.49 \vspace{1.0ex}\\ 5.39    
   \vspace{0.8ex}\end{array}$  
&  
$\begin{array}{c}   \displaystyle   
0.36 \vspace{1.0ex}\\ 6.46   
   \vspace{0.8ex}\end{array}$  
&  
$\begin{array}{c}   \displaystyle   
0.59 \vspace{1.0ex}\\ 7.56  
   \vspace{0.8ex}\end{array}$    
&  
$\begin{array}{c}   \displaystyle   
0.53 \vspace{1.0ex}\\ 8.17  
   \vspace{0.8ex}\end{array}$   
\\
\hline \hline  
\end{tabular}
\vspace{1em}  
%
%
%
%
\caption{\label{0.3,0.4,P,CC}%
Effective branching ratios for  $\,B\to\rho\pi\,$  decays,  
as in Table~\protect{\ref{0.3-P-CC}}, 
but with the additional kinematical cut  
$\,|\cos\theta|>0.4\,$  as explained in the text.  
}  \centering   \footnotesize
\vskip 0.5\baselineskip
\begin{tabular}{@{\hspace{1ex}}c@{\hspace{1ex}}|  
|@{\hspace{1ex}}c@{\hspace{1ex}}|@{\hspace{1ex}}c@{\hspace{1ex}}
|@{\hspace{1ex}}c@{\hspace{1ex}}|@{\hspace{1ex}}c@{\hspace{1ex}}
|@{\hspace{1ex}}c@{\hspace{1ex}}|@{\hspace{1ex}}c@{\hspace{1ex}}
|@{\hspace{1ex}}c@{\hspace{1ex}}|@{\hspace{1ex}}c@{\hspace{1ex}}}
\hline \hline  
Decay mode  &  $\vphantom{\biggl|}\rho$  &  $\sigma$  &  $B^*$  &  
$\rho+B^*$  &  $\rho+B^*+B_0^{}$  &  $\rho+\sigma$  &  
$\rho+\sigma+B^*$  &  $\rho+\sigma+B^*+B_0^{}$    
\\ \hline && && && && \vspace{-2ex} \\   
$\begin{array}{rcl}   \displaystyle
\bar B^0 & \hspace{-0.5em} \rightarrow & \hspace{-0.5em} \rho^0\pi^0 
\vspace{1.0ex}\\
B^- & \hspace{-0.5em} \rightarrow & \hspace{-0.5em} \rho^0\pi^- 
 \vspace{0.8ex}\end{array}$   
&  
$\begin{array}{c}   \displaystyle   
0.84 \vspace{1.0ex}\\ 3.81  \vspace{0.8ex}\end{array}$  
&  
$\begin{array}{c}   \displaystyle   
0.039 \vspace{1.0ex}\\ 4.79  \vspace{0.8ex}\end{array}$  
&  
$\begin{array}{c}   \displaystyle   
0.27 \vspace{1.0ex}\\ 1.71  \vspace{0.8ex}\end{array}$  
&  
$\begin{array}{c}   \displaystyle   
1.22\vspace{1.0ex}\\ 5.92 
    \vspace{0.8ex}\end{array}$    
&  
$\begin{array}{c}   \displaystyle   
1.11\vspace{1.0ex}\\ 6.58 
   \vspace{0.8ex}\end{array}$    
&  
$\begin{array}{c}   \displaystyle   
0.87 \vspace{1.0ex}\\ 7.97 
   \vspace{0.8ex}\end{array}$  
&  
$\begin{array}{c}   \displaystyle   
1.29 \vspace{1.0ex}\\ 8.63 
   \vspace{0.8ex}\end{array}$  
&  
$\begin{array}{c}   \displaystyle   
1.18 \vspace{1.0ex}\\ 9.13    
   \vspace{0.8ex}\end{array}$
\\ \hline \hline  
\end{tabular}
\vspace{1em} \\   
\end{table}

Finally, we must discuss a tacit assumption we have made
in the estimation of the $B^*$ and $B_0^{}$ contributions, 
which is made in Ref.~\cite{dea&co} as well. 
That is, in realizing the diagrams of Fig.~\ref{diagrams}(b), we have 
treated the strong $\bigl(B^*,B_0^{}\bigr) B\pi$ and weak   
$\,\bigl(B^*,B_0^{}\bigr)\to \pi\pi\,$  vertices as if the $B^*$ 
and $B_0^{}$  mesons were on their mass shell. 
This assumption is compatible with the assumed use of 
the combined heavy-quark and chiral limits in the 
treatment of the strong  $\bigl(B^*,B_0^{}\bigr)B\pi$  vertices. 
However, neither assumption is appropriate for  $\,B\to\rho\pi\,$  decay. 
That is, for the  $\,B\to3\pi\,$  decays 
of interest, we require that two of the three pions have 
an invariant mass $\sqrt s$ comparable to that of the $\rho$ meson.  
This implies that in most of the relevant phase-space region 
the mediating heavy-mesons carry $s$ values much smaller than 
their squared masses --- they are highly off-mass-shell.   
Moreover, the bachelor $\pi$ is never soft in this kinematical 
region.  
Thus the combined heavy-quark and chiral limits are used beyond 
their range of validity. 
These effects modify the vertices we have assumed in 
Eqs.~(\ref{BsBp}), (\ref{B0Bp}) and Eq.~(\ref{B*,B0->p}). 
Unfortunately, the needed off-shell extrapolations cannot
be done reliably, although we would generically 
expect this effect to suppress the numerical importance
of the $B^*$ and $B_0^{}$ contributions.   
For example, the form factors of Eqs.~(\ref{B*,B0->p}) now 
depend on both $q^2$ and $k^2$; the vertices are only 
``half'' off-shell,  so that $p^2$ does not enter, 
as the final-state $\pi$ is on its mass shell.   
Moreover, additional form factors appear.   
To illustrate, we note that the general parametrization 
\begin{eqnarray}
\bigl\langle \pi^+(p) \bigr| 
\bar u \gamma^\mu L b \bigl| \bar B_0^0(k) \bigr\rangle  
&=&  -{\rm i} F_0^{B_0^{}\to\pi}\bigl(k^2,p^2,q^2\bigr)
\frac{\left( M_{B_0^{}}^2-M_\pi^2 \right) q^\mu}{q^2}  
\nonumber \\
&&  -\,\,  
{\rm i}  F_1^{B_0^{}\to\pi} \bigl(k^2,p^2,q^2\bigr)
\left[ p^\mu + k^\mu - 
\left( M_{B_0^{}}^2-M_\pi^2 \right) \frac{q^\mu}{q^2} \right]
\end{eqnarray}
predicated by an assumption of Lorentz invariance yields 
\begin{eqnarray}   \label{hosB*,B0->p}    
q^\mu \bigl\langle \pi^+(p) \bigr| 
\bar u \gamma^\mu L b \bigl| \bar B_0^0(k) \bigr\rangle  
\,=\, 
-{\rm i} \left( M_{B_0^{}}^2-M_\pi^2 \right)  
F_0^{B_0^{}\to\pi} \bigl(q^2,k^2\bigr)   
+ {\rm i} \left( M_{B_0^{}}^2-k^2 \right)  
 F_1^{B_0^{}\to\pi} \bigl(q^2,k^2\bigr)  
\end{eqnarray}   
for the half-off-shell matrix element of interest. 
The matrix element is a linear combination of signed, 
uncertain contributions, so that its sign is ultimately unclear. 
Similar considerations apply to the  $\,B^*\to\pi\,$  matrix 
element, as well as to the strong vertices of
Eqs.~(\ref{BsBp}), (\ref{B0Bp}).   
In the treatment of Ref.~\cite{DesEHT}, an off-shell extrapolation 
of Eq.~(\ref{BsBp}), in the kinematic region of interest, 
is effected through the replacement 
$\sqrt{M_B M_{B^*}} \to \sqrt{M_B \sqrt{s}}$.
To assess the impact of these considerations on the numerical
results we have reported, we shall adopt a similarly {\it ad hoc} 
prescription. 
Thus, we perform the replacement
\begin{eqnarray}   
M_{B^*}^{3/2}  \,\,\to\,\,  s^{3/4}   \,\,   
\end{eqnarray}     
in the numerator of the  $B^*$ amplitudes in  Eq.~(\ref{A_B*}), 
so that the ``off-shellness'' of both the strong and weak 
vertices is taken into account. 
We neglect the $B_0$ in this simple numerical
estimate, as its effect was rather small to start with. 
We calculate the corresponding branching 
ratios and collect the results in  Table~\ref{0.3,OS2,P,CC}. 
Our simple prescription leads to a dramatic reduction of 
the  $B^*$  contributions, as a comparison with
Table~\ref{0.3-P-CC} makes clear. Note that the computed 
values of ${\cal R}$ are still consistent with the empirical
ones, as a reduction in ${\cal R}$ is still realized through
the $\sigma$ contributions. 
Although we cannot draw firm conclusions from this simple 
exercise, it serves to illustrate that neglecting   
the off-shell nature of the heavy-meson vertices in 
the kinematic region of interest could easily lead 
to a considerable overestimate of their effects.

\begin{table}[ht]   
\caption{\label{0.3,OS2,P,CC}%
Effective branching ratios for  $\,B\to\rho\pi\,$  decays,  
as in Table~\ref{0.3-P-CC}, except that the off-shellness 
of the $B^*$  meson is included as explained in the text. 
}  \centering   \footnotesize
\vskip 0.5\baselineskip
\begin{tabular}{@{\hspace{1ex}}c@{\hspace{1ex}}|  
|@{\hspace{1ex}}c@{\hspace{1ex}}|@{\hspace{1ex}}c@{\hspace{1ex}}
|@{\hspace{1ex}}c@{\hspace{1ex}}|@{\hspace{1ex}}c@{\hspace{1ex}}
|@{\hspace{1ex}}c@{\hspace{1ex}}|@{\hspace{1ex}}c@{\hspace{1ex}}}
\hline \hline  
Decay mode  &  $\vphantom{\biggl|}\rho$  &  $\sigma$  &  $B^*$  & 
$\rho+B^*$  &  $\rho+\sigma$  &  $\rho+\sigma+B^*$  
\\ \hline && && && \vspace{-2ex} \\   
$\begin{array}{rcl}   \displaystyle
\bar B^0 & \hspace{-0.5em} \rightarrow & \hspace{-0.5em} \rho^-\pi^+ \\     
\bar B^0 & \hspace{-0.5em} \rightarrow & \hspace{-0.5em} \rho^+\pi^- \\     
\bar B^0 & \hspace{-0.5em} \rightarrow & \hspace{-0.5em} \rho^0\pi^0 
\vspace{1.0ex}\\ 
B^-  & \hspace{-0.5em} \rightarrow & \hspace{-0.5em} \rho^0\pi^-  
  \vspace{0.8ex}\end{array}$   
&  
$\begin{array}{c}   \displaystyle   
16.0  \\  4.76 \\  0.86 \vspace{1.0ex}\\ 4.06 
  \vspace{0.8ex}\end{array}$  
&  
$\begin{array}{c}   \displaystyle   
0.001  \\  0.001 \\  0.065 \vspace{1.0ex}\\ 7.66 
  \vspace{0.8ex}\end{array}$  
&  
$\begin{array}{c}   \displaystyle   
0.03  \\  0.01 \\ 0.02 \vspace{1.0ex}\\ 0.25
  \vspace{0.8ex}\end{array}$  
&  
$\begin{array}{c}   \displaystyle   
16.0 \\ 4.85 \\ 0.88 \vspace{1.0ex}\\ 4.43
  \vspace{0.8ex}\end{array}$  
&  
$\begin{array}{c}   \displaystyle   
15.9 \\ 4.80 \\ 0.91 \vspace{1.0ex}\\ 11.1 
  \vspace{0.8ex}\end{array}$  
&  
$\begin{array}{c}   \displaystyle   
16.0 \\ 4.88 \\ 0.95 \vspace{1.0ex}\\ 10.7 
  \vspace{0.8ex}\end{array}$  
\\ \hline \hline && && && \vspace{-2ex} \\   
$\vphantom{\Bigl|}\cal R$
&  
5.1  
&  
-  
&  
-
&  
4.7  
&  
1.9  
&  
1.9  
\\  \hline \hline  
\end{tabular}
\vspace{1em} \\  
\end{table}

\section{Conclusions\label{conclusion}}    
   
We have examined resonant and nonresonant backgrounds 
to $\,B\to\rho\pi\to3\pi\,$  decays which can potentially
impact the extraction of $\alpha$ from a Dalitz-plot analysis
of $B\to \pi^+\pi^-\pi^0$ decays~\cite{qui&co}, as well as 
the value of the ratio of branching ratios we 
term ${\cal R}$, as defined in Eq.~(\ref{R_x}). 
In particular, we have evaluated the effects of nonresonant
contributions mediated by the heavy mesons  
$B^*$  and  $B_0^{}$, as well as the contributions from     
the light $\sigma$  resonance via $B\to\sigma\pi\to 3\pi$ decay, 
in the $\rho$-mass region. 
In this, our analysis parallels that of Refs.~\cite{dea&co,DeaPol}, 
though it differs fundamentally in two points. Firstly, 
we use the vector and scalar form-factors of
Ref.~\cite{GarMei}, which are 
consistent with low-energy theoretical constraints and thus are 
suitable for the description of broad resonant structures
such as the $\rho$ and the $\sigma$. 
The scalar form factor, in particular, is quite different 
from the Breit-Wigner form adopted in other analyses~\cite{DeaPol,e791}
and leads to differing results~\cite{GarMei}. Secondly, in the
kinematics of interest, the $B^*$ and $B_0^{}$ are highly 
off-mass-shell, impacting the strong and weak vertices which
mediate the  $\,B\to (B^*,B_0)\pi\to \pi\pi\pi\,$ decay.  
We find that these effects can reduce the heavy-meson 
contributions substantially.

Our numerical results show, were we to neglect the off-shell effects
we have mentioned, that the $B\to \rho^\pm\pi^\mp$  decay modes 
are little affected by the 
$\sigma$ and heavy-meson backgrounds, whereas the $\rho^0\pi^0$ 
mode receives large contributions from the latter.  
In contrast, the $B^-\to\pi^+\pi^-\pi^-$ decay mode 
contains large contributions 
from both the $\sigma$ and $B^*$, though the $\sigma$
contributions numerically dominate. Effecting a simple model
of off-shell effects, we find that the $B^*$ effects
are substantially reduced. The off-shell
extrapolation of interest cannot be effected with certainty; nevertheless,
our estimates indicate that the neglect of this effect may lead
to a substantial overestimate of the $B^*$ contributions in 
$B\to 3\pi$ decay. The role of the $\sigma$ in lowering the theoretical
value of ${\cal R}$ and yielding a favorable comparison with 
experiment persists despite these considerations. 
  
\bigskip 
 
{\it Note added.}\,\,  
Since the submission of this paper for publication, 
a report by the  BABAR Collaboration has appeared~\cite{B->3pi},
giving the experimental bound 
$\,{\cal B}\bigl(B^\pm\to\pi^+\pi^-\pi^\pm\bigr)  
< 15\times 10^{-6}\,$  at 90\%  C.L. 
This can be used to constrain the contribution of  
the  $B^*$-  and  $B_0^{}$-pole  diagrams. 
Using the  $g$ and $h$ values as in Table~\ref{0.3-P-CC}, we find   
$\,{\cal B}\bigl(B^-\to\pi^+\pi^-\pi^-)=24.8\times10^{-6}\,$   
for the combined  $\,\rho$+$\sigma$+$B^*$+$B_0^{}\,$  contribution,  
where we have integrated over all the allowed phase-space.    
Were we to use the intermediate values of $g$ and $h$ given in 
Table~\ref{0.3-med-P-CC}, though  such a $g$ is not favored by 
data~\cite{D*width}, we would obtain  
$\,{\cal B}\bigl(B^-\to\pi^+\pi^-\pi^-\bigr)=18.7\times10^{-6}.\,$    
If we use our off-shell extrapolation (neglecting the  
small $B_0^{}$ contribution) and the parameters of 
Table~\ref{0.3-P-CC}, we find 
$\,{\cal B}\bigl(B^-\to\pi^+\pi^-\pi^-\bigr)=15.4\times10^{-6}.\,$    
This comparison supports our assertion: the treatment of the  $B^*$  
vertices in  Ref.~\cite{dea&co} tends to yield an overestimate
of their contribution to  $\,B\to 3\pi\,$ decay.
On a related note, the failure to confront the empirical bound on 
$\,{\cal B}\bigl(B^-\to K^+ K^- \pi^-\bigr)\,$  decay has been
described in recent work by Cheng and Yang~\cite{CheYan}.

\bigskip
    
\noindent{\bf Acknowledgments}\,\,
We thank Ulf-G. Mei{\ss}ner for helpful discussions and 
J.A. Oller for the use of his scalar form factor program. 
The work of J.T. is supported by the U.S. Department of Energy 
under contract DE-FG01-00ER45832.  
S.G. thanks the SLAC Theory Group for hospitality and is supported by
the U.S. Department of Energy under contracts 
DE-FG02-96ER40989 and DE-AC03-76SF00515.


\end{document}